\documentclass[iop,apj,tighten]{emulateapj}

\shorttitle{LITHIUM ISOTOPE RATIO NEAR IC~443}
\shortauthors{TAYLOR ET AL.}

\begin{document}
\title{The $^7$Li/$^6$Li Isotope Ratio Near the Supernova Remnant IC~443\altaffilmark{1}}
\author{C. J. Taylor\altaffilmark{2}, A. M. Ritchey\altaffilmark{2}$^,$\altaffilmark{3}, S. R. Federman\altaffilmark{2}, and D. L. Lambert\altaffilmark{4}}
\altaffiltext{1}{Based on observations obtained with the Hobby-Eberly Telescope, which is a joint project of the University of Texas at Austin, the Pennsylvania State University, Stanford University, Ludwig-Maximilians-Universit\"at M\"unchen, and Georg-August-Universit\"at G\"ottingen.}
\altaffiltext{2}{Department of Physics and Astronomy, University of Toledo, Toledo, OH 43606, USA; corbin.taylor@rockets.utoledo.edu; steven.federman@utoledo.edu}
\altaffiltext{3}{Department of Astronomy, University of Washington, Seattle, WA 98195, USA; aritchey@astro.washington.edu}
\altaffiltext{4}{W. J. McDonald Observatory, University of Texas at Austin, Austin, TX 78712, USA; dll@astro.as.utexas.edu}

\begin{abstract}
We present an analysis of $^7$Li/$^6$Li isotope ratios along four sight lines that probe diffuse molecular gas near the supernova remnant IC~443. Recent gamma-ray observations have revealed the presence of shock-accelerated cosmic rays interacting with the molecular cloud surrounding the remnant. Our results indicate that the $^7$Li/$^6$Li ratio is lower in regions more strongly affected by these interactions, a sign of recent Li production by cosmic rays. We find that $^7$Li/$^6$Li~$\approx7$ toward HD~254755, which is located just outside the visible edge of IC~443, while $^7$Li/$^6$Li~$\approx3$ along the line of sight to HD~43582, which probes the interior region of the supernova remnant. No evidence of $^7$Li synthesis by neutrino-induced spallation is found in material presumably contaminated by the ejecta of a core-collapse supernova. The lack of a neutrino signature in the $^7$Li/$^6$Li ratios near IC~443 is consistent with recent models of Galactic chemical evolution, which suggest that the $\nu$-process plays only a minor role in Li production.
\end{abstract}


\section{INTRODUCTION}
The production of Li involves contributions from a number of different astrophysical sources, including the Big Bang, Galactic cosmic rays (GCRs), red giant branch (RGB) and asymptotic giant branch (AGB) stars, and Type~II supernovae (SNe II). The less abundant isotope, $^6$Li, is mainly a product of spallation (and $\alpha + \alpha$ fusion) reactions induced by GCRs in interstellar gas \citep[e.g.,][]{m71,r97}. Evidence for this can be seen in the solar system abundance ratio of $^6$Li to $^9$Be, the sole stable isotope of Be which can be produced only through GCR spallation. The meteoritic $^6$Li/$^9$Be ratio \citep[5.6;][]{l03} is essentially the GCR value. In contrast, the $^7$Li/$^6$Li ratio for solar system material is $\sim$12, while standard GCR nucleosynthesis predicts a ratio of $\sim$1.5. Since the primordial $^7$Li abundance arising from the Big Bang is approximately a factor of 10 less than the present-day abundance, the majority of $^7$Li must be produced in stars. However, the precise nature of the stellar source remains unclear.

The most promising candidates for a stellar $^7$Li source include RGB and AGB stars, where $^7$Li is produced via the Cameron-Fowler mechanism \citep[e.g.,][]{sb99}, and SNe II, where $^7$Li is synthesized by neutrino-induced spallation in the He and C shells of the progenitor star during core collapse \citep[i.e., the $\nu$-process;][]{w90}. Direct observational evidence for the $\nu$-process remains elusive, yet could be revealed by detailed studies of $^7$Li/$^6$Li isotope ratios in interstellar gas surrounding supernova remnants (SNRs). Since virtually no $^6$Li is expected to be produced by neutrino nucleosynthesis \citep[e.g.,][]{ww95,y08}, interstellar material contaminated by SN~II ejecta should exhibit a $^7$Li/$^6$Li ratio that is enhanced over the ambient interstellar value. However, a reduced $^7$Li/$^6$Li ratio might also be anticipated near a supernova remnant because SNRs are thought to be the primary sources responsible for cosmic-ray acceleration. Interactions between cosmic rays, accelerated by a supernova shock, and any nearby interstellar gas will drive the $^7$Li/$^6$Li ratio toward the pure GCR value.

The well-studied SNR IC~443 provides an excellent opportunity to test which of these processes dominates the production of Li by SNe~II. Located at a distance of 1.5~kpc in the Gem OB1 association, IC~443 is an intermediate-age (10$-$30 kyr) core-collapse SNR known to be interacting with atomic and molecular gas in its vicinity. Evidence of the interaction includes observations of shocked H~\textsc{i} filaments \citep[e.g.,][]{bs86,l08} and shocked molecular clumps \citep[e.g.,][]{h86,d92,s05}, as well as the detection of OH (1720~MHz) masers coincident with the shocked molecular material \citep{c97,h06}. Many recent investigations of IC~443 have focused on the emission of gamma rays at GeV and TeV energies \citep[e.g.,][]{a09,t10,a10}. These studies strongly suggest a pionic origin for the gamma-ray emission, indicating that shock-accelerated cosmic rays (in addition to the shocks themselves) are interacting with the ambient molecular cloud.

\begin{figure}
\centering
\includegraphics[width=0.45\textwidth]{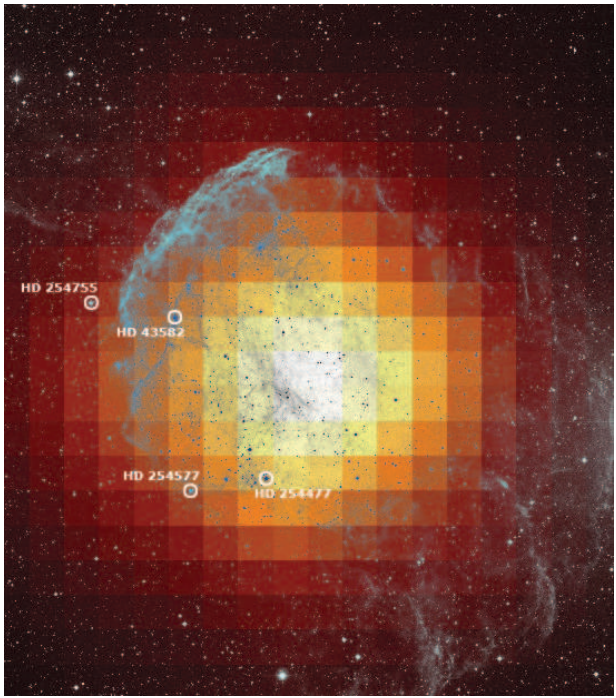}
\caption[]{Composite image of IC~443. The optical image from the Digitized Sky Survey (POSS-II/red filter) is shown superimposed onto a gamma-ray intensity map in the 400~MeV to 3~GeV energy range obtained by the \emph{Fermi} LAT (J. Hewitt 2011, private communication). The stars targeted for HET observations are labeled.}
\end{figure}

In this Letter, we present an analysis of $^7$Li/$^6$Li isotope ratios along four sight lines through IC~443. The ratios are extracted from high-resolution observations of Li~\textsc{i}~$\lambda6707$ toward HD~254477, HD~254577, HD~43582, and HD~254755. Figure~1 shows the positions of the target stars in relation to the gamma-ray emission detected by the \emph{Fermi} Large Area Telescope (LAT) in the 400~MeV to 3~GeV energy range (J. Hewitt 2011, private communication). Previous studies of the optical absorption profiles toward the selected targets have revealed extensive high-velocity gas \citep{ws03,h09}, indicating that these stars probe material directly associated with the SNR.

\section{OBSERVATIONS AND DATA REDUCTION}
The target stars were observed with the 9.2~m Hobby-Eberly Telescope (HET) at McDonald Observatory during the winters of 2008/2009 and 2009/2010. All observations employed the High Resolution Spectrograph \citep{t98} with an effective slit width of 125 $\mu$m ($R\approx120,000$) so that the fine structure lines of Li~\textsc{i} could be adequately resolved. Two spectrographic settings (centered at 4931 \AA{} and 5936 \AA{}) provided data on Li~\textsc{i}~$\lambda6707$. The setting at shorter wavelengths was needed to simultaneously obtain data on K~\textsc{i}~$\lambda4044$, while the longer wavelength setting allowed the K~\textsc{i}~$\lambda7698$ line to be observed. Both settings also yielded information on Ca~\textsc{i}~$\lambda4226$, CH$^+$~$\lambda4232$, and CH~$\lambda4300$.

Basic information concerning the target stars can be found in \citet{h09}. The three brighter targets were observed with both instrumental setups, resulting in total exposure times on Li~\textsc{i}~$\lambda6707$ of 6.1 hr for HD~254577 and 4.7 hr for HD~43582 and HD~254755. These exposure times yielded signal-to-noise ratios (S/N) per resolution element near Li~\textsc{i} of about 1100. The faintest star in our sample, HD~254477, was observed only with the longer wavelength setting for a total of 2.0 hr, resulting in a S/N of 380.

Standard procedures within IRAF were employed for bias correction, cosmic-ray removal, scattered light subtraction, and flat fielding. Ideally, a master flat was obtained from the median of flats taken for a given night. In some instances, only the first flat of a sequence could be used due to the appearance of an emission feature near 6708 \AA{} in subsequent flats. One-dimensional spectra were extracted from the processed images and were wavelength-calibrated after identifying emission lines in Th-Ar comparison spectra. Before shifting the calibrated spectra to the local standard of rest (LSR) frame, small velocity corrections were applied based on the measured wavelengths of the atmospheric [O~\textsc{i}] emission lines at 6300~\AA{} and 5577~\AA{}. This was necessary to account for offsets in the velocity zero points caused by slight changes in spectrograph placement between the stellar and Th-Ar exposures.

\begin{figure*}
\centering
\includegraphics[width=0.9\textwidth]{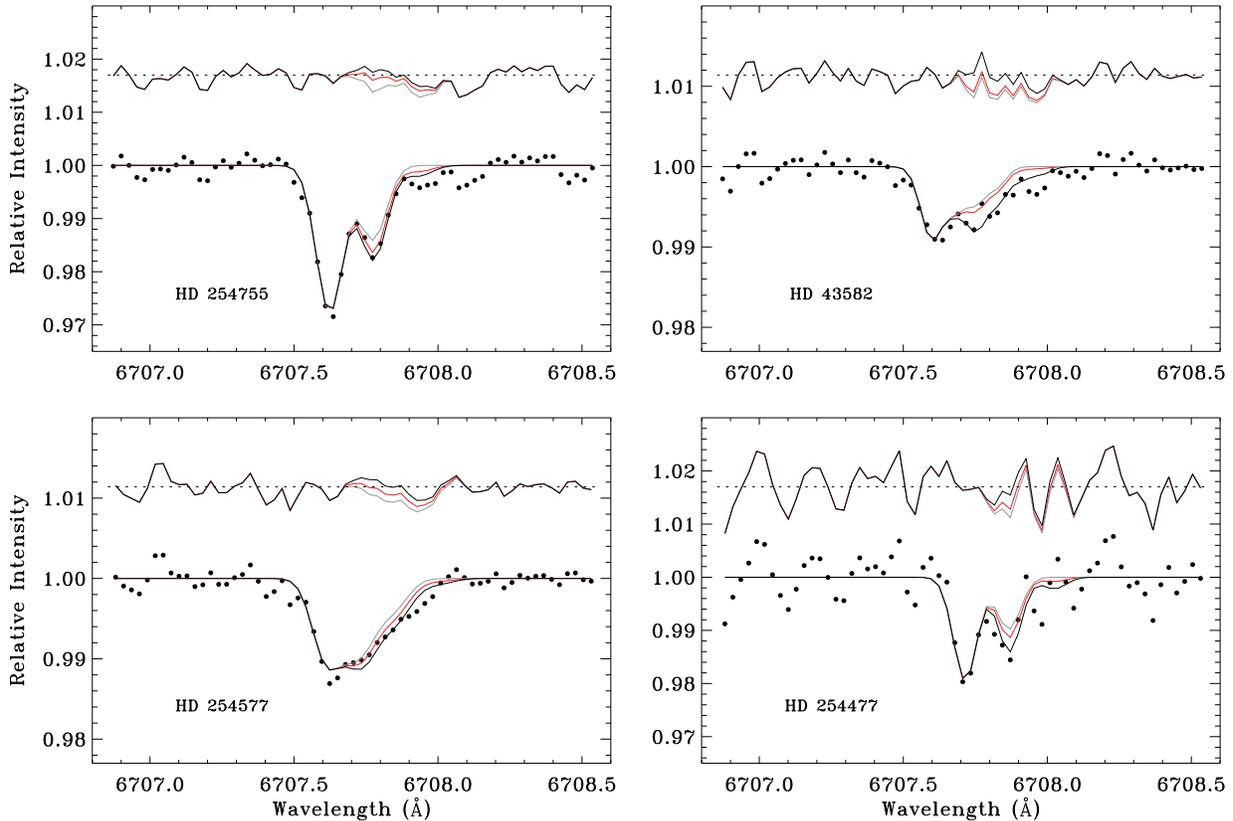}
\caption[]{Profile synthesis fits to the fine structure lines of $^7$Li~\textsc{i} and $^6$Li~\textsc{i} toward HD~254755, HD~43582, HD~254577, and HD~254477. Along with the best-fit synthetic profile (black line), two alternative syntheses are shown in each case, one assuming a solar system $^7$Li/$^6$Li ratio (red line) and one assuming no $^6$Li is present (grey line). The parameters for $^7$Li in these alternative syntheses are the same as in the best-fit cases. Residuals are given above each fit. For HD~43582 and HD~254577, the best-fit profile displayed is that based on the K~\textsc{i} template. Note that the alternative syntheses are less satisfactory in general, but particularly for HD~43582.}
\end{figure*}

Final spectra were produced by co-adding the individual exposures of a given target for orders containing the lines of interest. The co-added spectra were normalized to unity by fitting low-order polynomials to regions free of interstellar or telluric features. Figure~2 presents the reduced Li~\textsc{i} spectra for the four sight lines probing IC~443. From the widths of Th~\textsc{i} emission lines in the nightly comparison spectra, we found that the actual resolving power of the spectrograph during our observations was $R=98,000$.

\section{ANALYSIS}
Detailed knowledge of the velocity component structure in each direction is necessary if meaningful $^7$Li/$^6$Li ratios are to be extracted from the complicated Li~\textsc{i} line profiles, which exhibit fine and hyperfine structure in addition to isotopic splitting. For a single, optically-thin cloud, and in the absence of $^6$Li, the two fine-structure components of the Li~\textsc{i} doublet will exhibit relative strengths of 2:1 and a separation of 0.15 \AA. Addition of $^6$Li reduces this ratio because the stronger of the two $^6$Li components is superimposed onto the weaker of the components from $^7$Li (i.e., the isotope shift is approximately equal to the fine-structure separation).

Further complications arise when multiple clouds are present along the line of sight. Thus, the first step in our analysis was to derive a robust solution for the line-of-sight component structure using species with moderately strong absorption lines. The K~\textsc{i}~$\lambda7698$ and CH~$\lambda4300$ lines were of particular interest as these species are expected to coexist with Li~\textsc{i} in cool, diffuse clouds \citep{wh01,k03,p05}. While the weaker K~\textsc{i} line at 4044 \AA{} would provide a better template for Li~\textsc{i}~$\lambda6707$ (due to these lines having similar intrinsic strengths), absorption from K~\textsc{i}~$\lambda4044$ is detected only toward HD~254755. We also analyzed the Ca~\textsc{i}~$\lambda4226$ and CH$^+$~$\lambda4232$ absorption profiles to check for consistency in velocity among the components in common with K~\textsc{i} and CH.

The column densities and component parameters for all of the relevant species were determined through profile synthesis using the program ISMOD developed by Y. Sheffer \citep[see, e.g.,][]{s08}. ISMOD treats the velocities, $b$-values, and column densities of the absorption components as free parameters while minimizing the rms deviations in the residuals of the fit. Hyperfine structure is included in our synthetic profiles for Li~\textsc{i} and K~\textsc{i}, with wavelengths and $f$-values adopted from \citet{m03}. Our fits also account for $\Lambda$-doubling in the CH line. To extract the $^7$Li/$^6$Li ratio, the profile synthesis routine performs a simultaneous fit to the blended fine structure lines of $^7$Li~\textsc{i} and $^6$Li~\textsc{i}. The component structure is assumed to be identical for the two isotopes, with the isotope ratio left as a free parameter. Our results for K~\textsc{i} and CH were used to determine the number of velocity components to include in the synthesis for Li~\textsc{i}.

For HD~254755, a single component contributes more than 90\% of the total column density of K~\textsc{i} and CH. Our best one-component fit to the Li~\textsc{i} profile (Figure~2) yields a $^7$Li/$^6$Li ratio of $7.1\pm2.4$, where the errors are dominated by the observational uncertainties associated with the detection of $^6$Li. The velocity and $b$-value resulting from the fit (see Table~1) are in good agreement with those for the corresponding component in K~\textsc{i} ($v_{\mathrm{LSR}}=-6.4$~km~s$^{-1}$; $b=1.2$~km~s$^{-1}$) and CH ($v_{\mathrm{LSR}}=-5.9$~km~s$^{-1}$; $b=2.1$~km~s$^{-1}$).

For HD~43582 and HD~254577, two velocity components comprise 70$-$80\% of the total columns of K~\textsc{i} and CH. The Li~\textsc{i} profiles in these directions are manifestly more blended than toward HD~254755, indicating that multiple components are present. We therefore kept the component structure fixed in the Li~\textsc{i} syntheses, with component parameters determined separately from the two dominant components in K~\textsc{i} and CH. The results are given in Table~1 and details concerning the fits are presented in Figures~3 and 4 for HD~43582 and HD~254577, respectively. In both cases, the $^7$Li/$^6$Li ratios derived from the K~\textsc{i} and CH templates are indistinguishable given the uncertainties. Thus, we adopt the average ratios for further analysis (i.e., $^7$Li/$^6$Li~=~$3.1\pm1.4$ for HD~43582 and $6.1\pm3.2$ for HD~254577).

For HD~254477, the strongest velocity component accounts for only about 60\% of the K~\textsc{i} and CH column density, but no other component makes up more than 10\% of the profile in these species. Our best one-component fit to the Li~\textsc{i} profile yields a consistent velocity and $b$-value compared to the dominant component in K~\textsc{i} ($v_{\mathrm{LSR}}=-2.8$~km~s$^{-1}$; $b=2.3$~km~s$^{-1}$) and CH ($v_{\mathrm{LSR}}=-2.5$~km~s$^{-1}$; $b=1.0$~km~s$^{-1}$). However, the significantly lower S/N makes it impossible to confidently derive a $^7$Li/$^6$Li ratio in this direction. Nominally, we find $^7$Li/$^6$Li~=~$4.5\pm3.6$, but with $N$($^6$Li~\textsc{i})~=~$(2.2\pm1.8)\times10^9$~cm$^{-2}$ (i.e., not even a $2\sigma$ detection). A more appropriate $3\sigma$ upper limit on $N$($^6$Li~\textsc{i}) of $\lesssim5.4\times10^9$~cm$^{-2}$ gives a lower limit on $^7$Li/$^6$Li of $\gtrsim1.9$.

\begin{figure}[!t]
\centering
\includegraphics[width=0.45\textwidth]{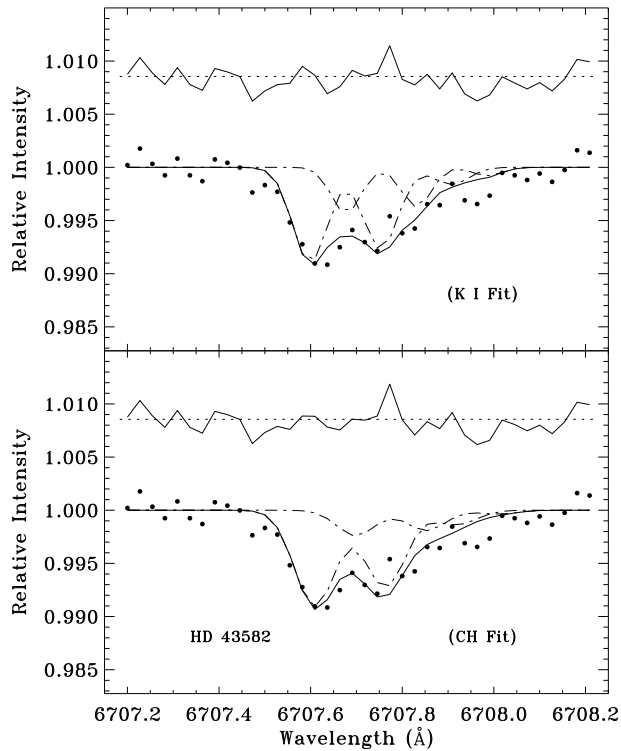}
\caption[]{Synthesis fits to the Li~\textsc{i} profile toward HD~43582, based on the K~\textsc{i} template (upper panel) and the CH template (lower panel). The dot-dashed lines show the contributions from individual cloud components to the overall profile, shown as a solid line.}
\end{figure}

\section{DISCUSSION AND CONCLUSIONS}
Our results on $^7$Li/$^6$Li isotope ratios in the vicinity of IC~443 shed new light on Li production associated with SNe~II. Before these observations, it was not clear whether one should expect an enhanced abundance of $^7$Li near SNe~II, resulting from neutrino spallation during the core-collapse phase, or a higher relative abundance of $^6$Li, due to spallation and fusion reactions initiated by shock-accelerated cosmic rays. Our analysis of Li~\textsc{i} absorption along four sight lines through IC~443 suggests that cosmic-ray interactions dominate the production of Li at least for this particular SNR.

All of the $^7$Li/$^6$Li ratios we obtain near IC~443 are lower than the solar system (meteoritic) value of 12.2 \citep{l03}, which \citet{k03} suggest is representative of gas in the solar neighborhood. While our results for HD~254577 and HD~254755 are actually consistent with the solar system ratio at the $2\sigma$ level, the ratio we find toward HD~43582 is lower by more than $6\sigma$. There is evidence, however, that the Li isotope ratio in the interstellar medium (ISM) in general is lower than the solar system value. The weighted mean ISM ratio from measurements available in the literature \citep{l93,m93,l95,k00,h02,k03,k09} is $7.3\pm0.6$ (if one considers only the strongest absorption component in each case). This value is very similar to the ratio we find toward HD~254755, which lies beyond the outer edge of the H$\alpha$ emission contours in the northeast of IC~443 (see Figure~1). The gas in this direction may thus represent material relatively unaffected by the supernova shock, implying that a $^7$Li/$^6$Li ratio of $\sim$7 characterized the ambient molecular cloud before the occurrence of the supernova.

\begin{figure}[!t]
\centering
\includegraphics[width=0.45\textwidth]{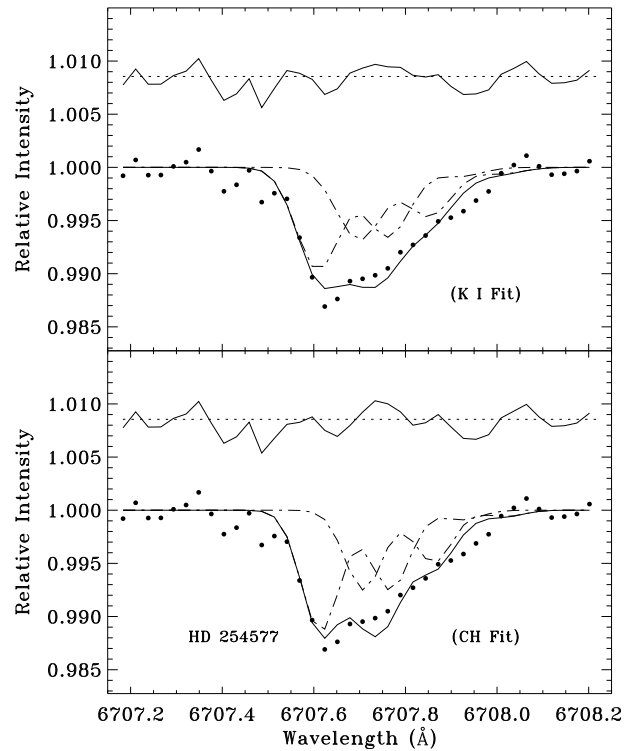}
\caption[]{Same as Figure~3 except for the Li~\textsc{i} profile toward HD~254577.}
\end{figure}

In contrast, the $^7$Li/$^6$Li ratio toward HD~43582 is 40$-$80\% lower than the mean value observed in the local ISM. The stars HD~254755 and HD~43582 are separated by only 7$^{\prime}$.4 on the sky (or 3.2 pc at the distance of IC~443), but the line of sight to HD~43582 clearly penetrates the interior region of the SNR. In fact, HD~43582 is the only star of those investigated by \citet{h09} that exhibits Ca~\textsc{ii} absorption at high positive velocity, strongly suggesting that it lies behind the receding edge of the expanding shell of the remnant. If the molecular gas in this direction has been subjected to a more intense cosmic-ray irradiation than that toward HD~254755, then an enhancement in the relative abundance of $^6$Li would be a natural consequence. Indeed, the $^7$Li/$^6$Li ratio toward HD~43582 approaches the value predicted by models of GCR nucleosynthesis \citep[$\sim$1.5;][]{m71,r97,l98}. The sight lines to HD~254577 and HD~254477 presumably also pass through the interior of the SNR. However, given the uncertainties, the $^7$Li/$^6$Li ratios in these directions are consistent with both the mean ISM value (as observed toward HD~254755) and the lower value we find toward HD~43582.

The discovery of a low $^7$Li/$^6$Li ratio in IC~443 corroborates the conclusions drawn from gamma-ray emission studies of the region \citep[e.g.,][]{a09,t10,a10}, which strongly suggest that the gamma radiation results from the decay of neutral pions produced through cosmic-ray interactions with molecular gas. Complementary results on the cosmic-ray ionization rate were recently reported by \citet{i10}. These authors find large H$_3^+$ column densities near IC~443 and deduce an H$_2$ ionization rate of $\zeta_2\approx2\times10^{-15}$~s$^{-1}$, or about five times the rate typically found for diffuse molecular clouds \citep[e.g.,][]{i07}.

Both the H$_3^+$ abundances discussed by \citet{i10} and the $^7$Li/$^6$Li ratios presented here indicate regions of enhanced cosmic-ray activity within IC~443. However, the two observables present potentially interesting differences in terms of which sight lines exhibit the largest enhancements. \citet{i10} infer an ionization rate of $\zeta_2=2.6^{+1.3}_{-1.9}\times10^{-15}$~s$^{-1}$ (the highest in their survey) toward HD~254577, yet report only an upper limit on $\zeta_2$ toward HD~43582 since H$_3^+$ is not detected. (The upper limit is $\lesssim1.4\times10^{-15}$~s$^{-1}$ if uncertainties in the H$_2$ column density are included.) These results appear to be in conflict with our determinations of $^7$Li/$^6$Li for these directions, since we find a lower $^7$Li/$^6$Li ratio toward HD~43582 than toward HD~254577. However, the uncertainties both in our determinations and in those of Indriolo et al., which rely on a number of assumptions concerning the physical conditions in the clouds containing H$_3^+$, are large. Moreover, since the abundance of H$_3^+$ traces the current cosmic-ray ionization rate, while the Li isotope ratio probes the integrated cosmic-ray flux, some differences in these two observables might be expected (B. Fields 2011, private communication).

A decrease in the $^7$Li/$^6$Li ratio due to enhanced cosmic-ray activity (resulting in newly-synthesized $^6$Li and $^7$Li) should be accompanied by an increase in the elemental Li abundance. If the clouds toward HD~43582 initially possessed a $^7$Li/$^6$Li ratio of $\sim$7, then the presently-observed ratio of 3.1 would imply a factor-of-two increase in the elemental Li abundance. Some evidence for this effect can be seen in the $N$(Li~\textsc{i})/$N$(K~\textsc{i}) ratios found toward the stars in our sample: $(2.7\pm0.2)\times10^{-3}$ for HD~254755, $(5.9\pm0.6)\times10^{-3}$ for HD~254577, $(6.4\pm0.9)\times10^{-3}$ for HD~43582, and $(6.8\pm1.4)\times10^{-3}$ for HD~254477. The three sight lines that pass through the interior of the remnant seem to exhibit the expected factor-of-two enhancement in the abundance of Li (relative to K) if the initial abundances were similar to that seen toward HD~254755.

Still, none of the $N$(Li~\textsc{i})/$N$(K~\textsc{i}) ratios we find near IC~443, are entirely unusual compared to values typical of diffuse clouds \citep[e.g.,][]{wh01,k03}. Without directly determining the elemental Li/K ratios, which would require making uncertain corrections for ionization and depletion onto grains, it is difficult to conclude whether an enhancement in Li is indeed observed. We can, however, state that a localized enhancement would be energetically feasible. If we assume that $n_{\mathrm{H}}\sim200$ cm$^{-3}$ for the gas toward HD~43582 \citep[as adopted by][]{i10}, then a factor-of-two increase in the Li abundance would require that $\sim$$3\times10^{50}$ erg are imparted to the accelerated cosmic rays by the SN shock \citep[see][]{r00}. This is only $\sim$20\% of the typical mechanical energy available in SN~II ejecta \citep{ww95}

Our observations suggest that the abundance of $^6$Li relative to $^7$Li has been enhanced near IC~443 by interactions between shock-accelerated cosmic rays and the ambient molecular cloud. At the same time, we find no evidence of $^7$Li synthesis by neutrino-induced spallation in material that presumably has been contaminated by the ejecta of a core-collapse supernova. Given the age of the remnant, the hot ejecta should have had ample time to cool, as it interacts with its dense surroundings, to the point at which absorption lines from neutral atoms would become visible. For comparison, \citet{w95} estimated a cooling time of 800 yr for the Vela remnant, which is about the same age as IC 443 but where the ambient ISM is of much lower density. The lack of a neutrino signature is consistent with recent models of Galactic chemical evolution \citep[e.g.,][]{p12}, which suggest a very minor role for the $\nu$-process in producing $^7$Li. Future measurements of $^7$Li/$^6$Li in gas surrounding other SNRs will help to establish unequivocally the role that neutrino spallation plays in Li production.

\acknowledgments
We gratefully acknowledge useful conversations with Brian Fields and Nick Indriolo. We also thank Jack Hewitt for providing us with the \emph{Fermi} data for Figure~1. C. Taylor participated in the Research Experience for Undergraduates program at the University of Toledo under NSF-REU grant PHY-1004649. D.L.L. acknowledges support from the Robert A. Welch Foundation of Houston, Texas through grant F-634. The Hobby-Eberly Telescope is a joint project of the University of Texas at Austin, the Pennsylvania State University, Stanford University, Ludwig-Maximilians-Universit\"at M\"unchen, and Georg-August-Universit\"at G\"ottingen. The HET is named in honor of its principal benefactors, William P. Hobby and Robert E. Eberly.

\clearpage

\begin{deluxetable}{lcccccc}
\tablecolumns{7}
\tablewidth{0pt}
\tablecaption{Li~\textsc{i} Profile Synthesis Results}
\tablehead{ \colhead{Star} & \colhead{Template} & \colhead{$v_{\mathrm{LSR}}$} & \colhead{$b$} & \colhead{$N$($^7$Li~\textsc{i})} & \colhead{$^7$Li/$^6$Li} & \colhead{$N$(Li~\textsc{i})} \\
\colhead{} & \colhead{} & \colhead{(km s$^{-1}$)} & \colhead{(km s$^{-1}$)} & \colhead{($10^9$ cm$^{-2}$)} & \colhead{} & \colhead{($10^9$ cm$^{-2}$)} }
\startdata
HD 254477 & \ldots & $-$2.2 & 1.7 & $10.2\pm1.8$ & $\gtrsim1.9$ & $12.4\pm2.5$\tablenotemark{a} \\
\\
HD 254577 & K~\textsc{i} & ($-$6.9) & (2.3) & $5.8\pm0.7$ & $5.7\pm3.1$ & $11.8\pm1.3$ \\
 & & ($-$2.9) & (2.4) & $4.2\pm0.7$ & \\
 & CH & ($-$6.7) & (1.8) & $6.1\pm0.6$ & $6.4\pm3.4$ & $11.6\pm1.1$ \\
 & & ($-$2.5) & (1.7) & $3.9\pm0.6$ & \\
\\
HD 43582 & K~\textsc{i} & ($-$7.4) & (1.4) & $4.3\pm0.6$ & $2.8\pm1.0$ & $8.3\pm1.1$ \\
 & & ($-$3.9) & (0.6) & $1.8\pm0.5$ & \\
 & CH & ($-$6.9) & (2.0) & $5.1\pm0.7$ & $3.3\pm1.7$ & $8.3\pm1.3$ \\
 & & ($-$2.9) & (1.8) & $1.3\pm0.6$ & \\
\\
HD 254755 & \ldots & $-$6.3 & 2.2 & $16.6\pm0.8$ & $7.1\pm2.4$ & $18.9\pm1.1$ \\
\enddata
\tablecomments{Velocities and $b$-values shown in parentheses were held fixed during profile synthesis. The last column gives the total Li~\textsc{i} column density. The corresponding K~\textsc{i} column densities for components also seen in Li~\textsc{i} are $18.3\pm0.1$, $19.7\pm0.1$, $12.9\pm0.1$, and $69.4\pm1.7$ (in units of $10^{11}$~cm$^{-2}$) for HD~254477, HD~254577, HD~43582, and HD~254755, respectively.}
\tablenotetext{a}{Assumes $^7$Li/$^6$Li~$\sim4.5$.}
\end{deluxetable}

\end{document}